\documentclass[a4paper]{article}

\usepackage{INTERSPEECH2021}

\title{Speak Like a Professional: Increasing Speech Intelligibility by Mimicking Professional Announcer Voice with Voice Conversion}
\name{Tuan Vu Ho$^1$, Maori Kobayashi$^2$, Masato Akagi $^3$}
\address{
  $^{1,3}$ School of Information Science, Japan Advanced Institute of Science and Technology \\
  $^2$ School of Human Sciences,Waseda University}
\email{tuanvu.ho@jaist.ac.jp, maori-k@aoni.waseda.jp, akagi@jaist.ac.jp}

\begin{document}

\maketitle
\begin{abstract}
In most of practical scenarios, the announcement system must deliver speech messages in a noisy environment, in which the background noise cannot be cancelled out. The local noise reduces speech intelligibility and increases listening effort of the listener, hence hamper the effectiveness of announcement system. There has been reported that voices of professional announcers are clearer and more comprehensive than that of non-expert speakers in noisy environment. This finding suggests that the speech intelligibility might be related to the speaking style of professional announcer, which can be adapted using voice conversion method. Motivated by this idea, this paper proposes a speech intelligibility enhancement in noisy environment by applying voice conversion method on non-professional voice. We discovered that the professional announcers and non-professional speakers are clusterized into different clusters on the speaker embedding plane. This implies that the speech intelligibility can be controlled as an independent feature of speaker individuality. To examine the advantage of converted voice in noisy environment, we experimented using test words masked in pink noise at different SNR levels. The results of objective and subjective evaluations confirm that the speech intelligibility of converted voice is higher than that of original voice in low SNR conditions.
\end{abstract}

\section{Introduction}
In most of the situation, public announcement systems must deliver speech messages through adverse listening environments with various competing sounds and reverberation. These negative conditions degrade intelligibility of the announcement speech, and in some cases, damage the integrity of the intended message. One solution to maintain the speech intelligibility is to increase the intensity of playback volume to increase the signal-to-noise ratio (SNR). However, this approach is only useful to some degrees due to the limitation of playback equipment power and the comfort of listener. Consequently, various approaches have been proposed to increase the speech intelligibility in noisy environments without increasing the total power of speech. These include modification of spectral properties \cite{Kleijn2015OptimizingSI, Taal2013SIIbasedSP, Taal2014SpeechER}, dynamic range compression \cite{Zorila2014OnSA, Zorila2012SpeechinnoiseII, Chermaz2020ASE}, modification of speech modulation spectrum \cite{AmanoKusumoto2005ModulationEO, Ngo2021IncreasingSI}, and time-scale modification \cite{Tang2011SubjectiveAO, Aubanel2013InformationpreservingTR}.

It is known that the voice-related professions, such as professional announcers, voice actor, and singer, can produce speech with the impression of clearer and easier to hear than normal person \cite{Noh2012HowDS, Kashimada2009EffectsOV}. Moreover, recent studies have shown that the speech from professional announcer can maintain its intelligibility better than speech from non-expert person in very noisy environment \cite{Kobayashi2020Intel}. This phenomenon can be exploited to inspire speech enhancement algorithms that seek to improve speech intelligibility in noise.
In this work, we aim to apply voice conversion technique to mimic a speaking style of the professional announcer. Voice conversion refers to the process of modifying voice personality without changing the linguistic information conveyed in speech waveform. Furthermore, voice conversion can be applied to control different attributes of voice style, i.e., gender and accent \cite{Ho2019NonparallelVC}. This study will clarify whether the announcer-adapted speech from voice conversion model still inherits the noise-resistance property of natural announcer speech. In addition, this study proposes a method to increase the speech intelligibility without completely change the voice individuality, which is useful in the situation where the identity of the speaker needs to be preserved. 

The structure of voice conversion model and the training procedure are described in Section 2. Next, the detail for experimental settings is shown in Section 3. Then the results of objective and subjective evaluations are described in Section 4. Finally, we conclude our paper Section 5 with some discussions for the future work. 

\section{StarGAN-v2 voice conversion model}
\begin{figure*}
    \centering
    \includegraphics[width=0.9\linewidth]{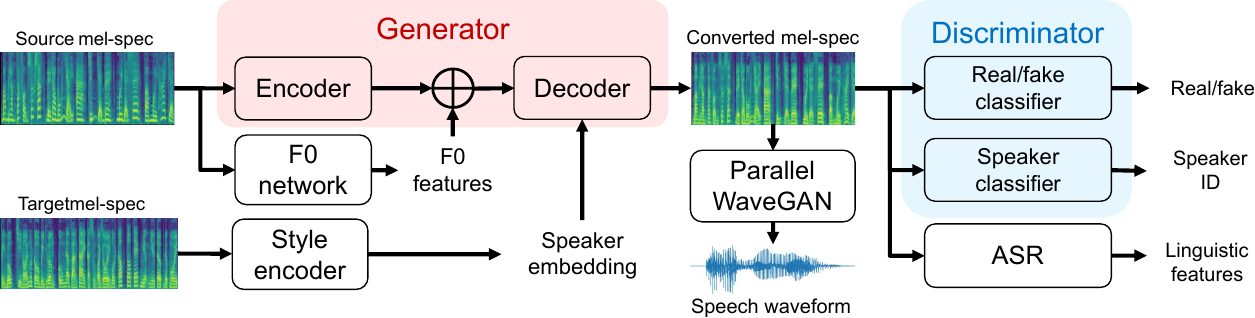}
    \caption{Overview structure of StarGAN-v2 voice conversion model. The model consists of a generator network to convert input mel-spectrogram, a discriminator network for adversarial training, a style encoder to extract speaker embedding, a pretrained F0 network for extracting F0 feature, a pretrained speech recognition to extract linguistic features, and a pretrained Parallel WaveGAN vocoder to generate waveform from mel-spectrogram.}
    \label{fig:stargan}
\end{figure*}
In this paper, a voice conversion method based on StarGANv2 \cite{Choi2020StarGANVD} model is used. We follow the official implementation of StarGANv2-VC \cite{Li2021StarGANv2VCAD} which is publicly available at \footnote{https://github.com/yl4579/StarGANv2-VC}. The overview structure of voice conversion model is depicted in Fig. \ref{fig:stargan}. The models consist of 6 modules as follows:
\begin{itemize}
    \item Style encoder network consists of a stack of 4 residual network (resnet) \cite{He2016DeepRL} layers an average pooling layer to extract the speaker embedding $s_{\textrm{emb}}$ from mel-spectrogram. The dimension of speaker embedding vector is set to 128.
    \item F0 network is pretrained to predict the F0 value and voice/un-voice region from the input mel-spectrogram. The F0 network is a stack of 2 convolutional layers, 3 resnet layers and a recurrent bidirectional long-short term memory layer. The output of the last resnet layer is used as the F0 features for generator network.
    \item Speech recognition (ASR) network is pretrained to predict the phoneme sequence from mel-spectrogram. The ASR network is a joint CTC-attention VGG-BLSTM network given by the Espnet Toolkit \footnote{https://github.com/espnet/espnet} \cite{Watanabe2018ESPnetES}.
    \item Parallel WaveGAN \cite{Yamamoto2020ParallelWA} vocoder is pretrained to generate speech waveform from input mel-spectrogram. The pretrained checkpoint is publicly available at \footnote{https://github.com/kan-bayashi/ParallelWaveGAN}.
    \item The generator network (G) consists of two sub-modules: an encoder network and a decoder network. The encoder takes the input mel-spectrogram and generate the hidden feature vector. The concatenation of hidden feature vector and F0 feature vector is fed to the decoder network to generate converted mel-spectrogram conditioned on the input speaker embedding.
    \item Discriminator network consists of two sub-modules: a real-fake classifier trained to discriminate real and converted mel-spectrogram, and a classifier to predict the speaker identity of input mel-spectrogram.
\end{itemize}

\subsection{Training procedure}
The training data consists of utterances from 20 professional announcers from ATR \cite{ATRdataset} dataset A-set and 20 non-expert speakers from ATR dataset C-set. All the utterances are preprocessed by resampling to 24 kHz, removing leading and trailing silence, and combining to 5-second chunks. There is total 22.234 utterances, in which 500 utterances are used for validation. The 80-band log-mel spectrogram with band limited frequency range (0 to 8 kHz) is extracted using short-time Fourier transform. The window length and frame shift are set to 1024 and 256 respectively.
We follow training strategy as described in \cite{Li2021StarGANv2VCAD} with the same objective functions and hyper-parameters. The voice conversion model is trained for 50 epochs with batch size of 48 using 2 Nvidia RTX3090 GPUs. The training process takes approximately 1 day to finish. 

\subsection{Visualization of speaker embedding}
\begin{figure}[bt]
    \centering
    \includegraphics[width=0.42\textwidth]{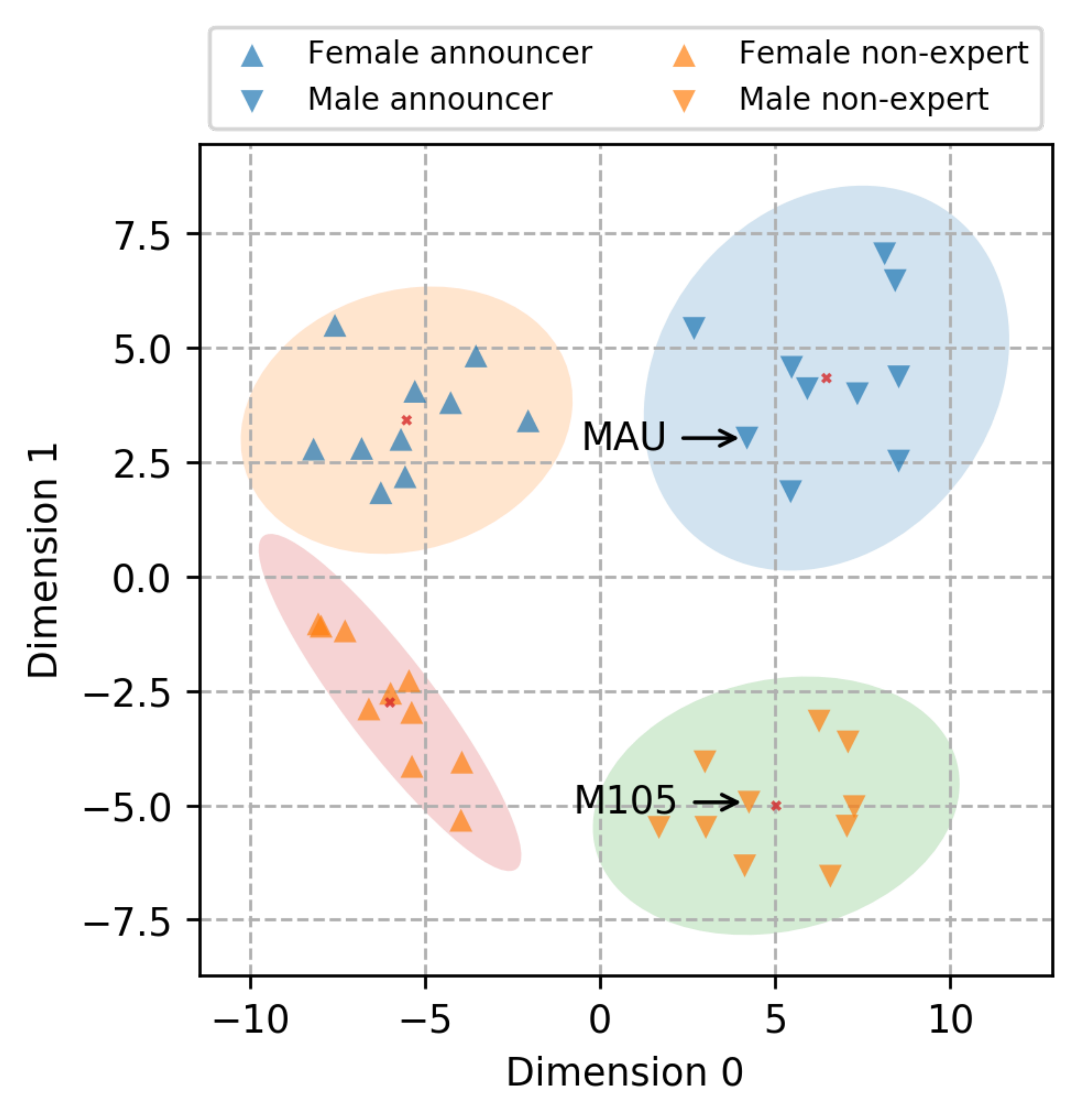}
    \caption{2D visualization of first and second principal components of speaker embedding. The red dots and eclipse shades denote the centroid and covariance of each cluster.}
    \label{fig:speaker_emb}
\end{figure}
The speaker embedding is a vector that encodes speaker individuality conveyed in the input mel spectrogram into a compact vector. By analyzing the speaker embedding using principal component analysis (PCA), we can factorize out the dominant features of speaker individuality. Figure 2 plots the first and second principal components of the speaker embedding after training. The first component corresponds to the gender of the speakers, while the second component corresponds to the voice type, which is non-expert voice or professional announcer voice. This result suggests that the style of voice, i.e., non-expert style or announcer style, can be controlled independently from other voice attributes.

\section{Experimental settings}
\subsection{Approach 1}
From the study of Kobayashi et al. [13], it has been shown that speech from professional announcers is perceptually clearer and easier to hear than that of non-expert voice even in noisy environments. This phenomenon is mainly originated from the differences of voice quality, clarity, projection, ring or resonant. However, it is unclear that whether the announcer-adapted voice can still possess this property of natural announcer voice. To investigate this point, we transform speaker individuality of speaker M105 (non-expert in ATR C-set) to that of target speaker MAU (professional announcer in ATR A-set) using voice conversion model. The speaker embedding of MAU speaker is used to synthesize the converted stimuli.
\subsection{Approach 2}
As can been clearly seen in Fig. \ref{fig:speaker_emb}, the second principal component of the speaker embedding captures the difference between non-expert speaker and professional announcer. This behavior is advantageous as the non-expert voice can be converted to have the voice style of professional announcers only by changing the second principal components of speaker embedding. Since the professional announcers are more intelligible even in noisy environment, it is expected that the second principal component can be used to increase the intelligibility of non-expert voice. To clarify this point, we propose to replace the second principal component of M105 speaker embeddings with the average value calculated from the second principal component of all male professional announcers. Then, the obtained speaker embedding is used to synthesize converted stimuli. Different from Approach 1, the speaker individuality of source speaker is partly changed as other principal components of speaker embeddings are preserved.

\subsection{Experiment speech stimuli}
We select 520 Japanese words, each may contain 1 to 4 mora, from the ATR Digital Voice Database A-set (ATR-A) and ATR Digital Voice Database C-set (ATR-C) as the clean stimuli for target and source speaker. All speech waveforms are resampled to 16 kHz sampling rate. There are 4 types of speech stimuli in the experiments, which are denoted as follows:
\begin{itemize}
    \item \textbf{Non-expert}: Natural speech of non-professional speaker, which is collected from speaker M105 in ATR-C set.
    \item \textbf{Announcer}: Natural speech of professional announcer, which is collected from speaker MAU in ATR-A set.
    \item \textbf{VC-1}: converted speech from speaker M105 to speaker MAU (Approach 1) by voice conversion model.
    \item \textbf{VC-2}: converted speech by shifting the second principal components of speaker embedding of speaker M105 (Approach 2) by voice conversion model.
\end{itemize}

To create the noisy stimuli, we mask the clean stimuli with pink noise at 5 different SNR levels: -9dB, -6dB, -3dB, 0dB, and $\infty$ (no noise). We calculate the root-mean-square of the speech signal only in the voice region and scale the noise signal to match with the desired SNR level. The voice region is derived from the text transcription of ATR dataset. To avoid the effect of different onset and offset timing between speech stimuli, duration of each stimulus is adjusted to contain the same 200ms of leading noise and 200ms of trailing noise. In addition, speech stimuli are gated with two raised cosine onset and offset windows of 40-ms to avoid overshoot distortion.

\section{Evaluations}
\subsection{Objective evaluation}

\begin{figure}[bt]
    \centering
    \includegraphics[width=0.45\textwidth]{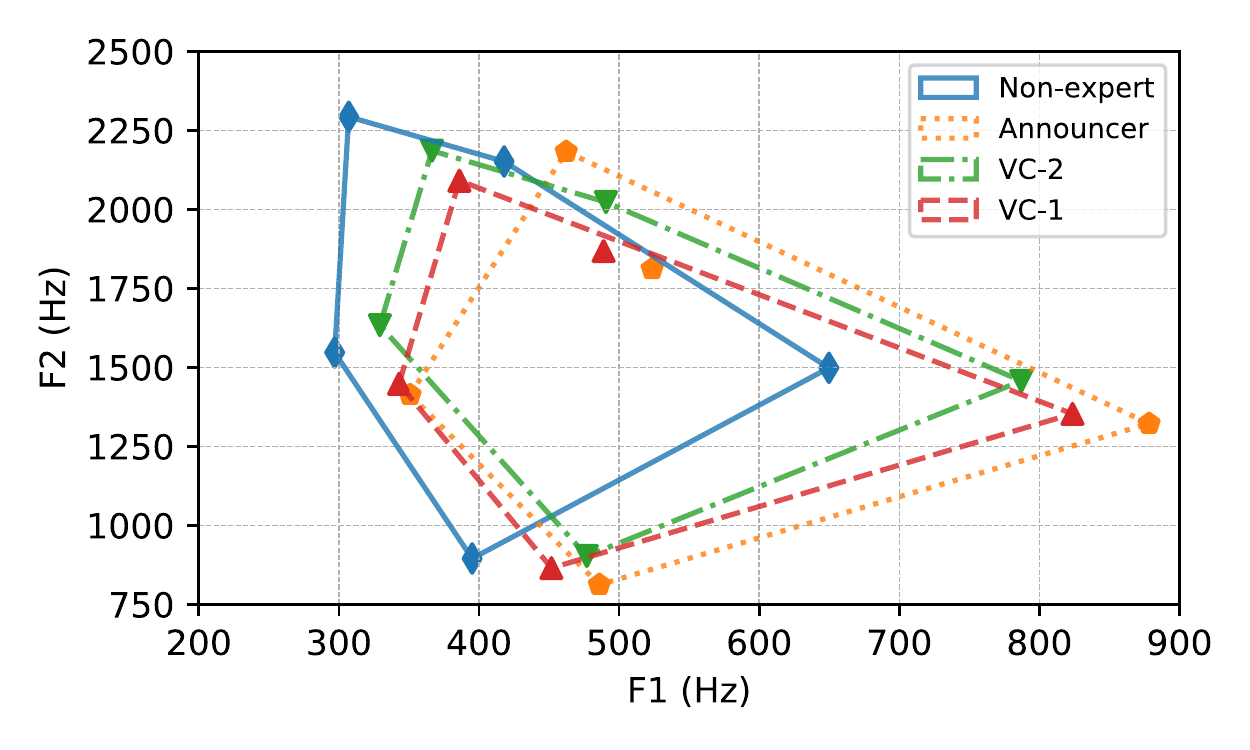}
    \caption{Average vowel space of non-expert, announcer, VC-1, and VC-2 stimuli, with respective areas: $2.64$, $3.60$, $2.91$ and $2.95$ ($\times10^5$ $\textrm{Hz}^2$).}
    \label{fig:vowel_space}
\end{figure}
Two objective metrics are used to evaluate the intelligibility of converted speech: 1) average vowel space, and 2) extended short-time objective intelligibility (eSTOI) \cite{Taal2010ASO}.
\subsubsection{Vowel space analysis}
Several studies have reported that the expansion of vowel space corresponds to an increase of speech intelligibility \cite{AmanoKusumoto2011ARO, Godoy2014ApproachingSI}. Based on this information, we compare the areas of the average vowel spaces derived from different types of stimuli. The formant frequencies of 5 Japanese vowels (/a/, /e/, /i/, /o/, and /u/) are extracted using Praat tool \cite{Boersma2002PraatAS}. The locations of vowels in each utterance are determined using the provided text transcription. Then the average frequency of first and second formants of vowels are calculated across all speech utterances. The vowel space is defined as the smallest polygon that fits all the vowels.
As can be seen from Fig. \ref{fig:vowel_space} the average vowel space of professional announcer exhibits the largest area ($3.6\times10^5$ $\textrm{Hz}^2$). This result aligns with the above assumption that the speech intelligibility increases with a larger vowel space. Interestingly, the converted voices from non-expert speaker (VC-1 and VC-2) show an expansion of vowel space from the non-expert vowel space, from $2.64\times10^5$ $\textrm{Hz}^2$ to $2.91\times10^5$ $\textrm{Hz}^2$ and $2.95\times10^5$ $\textrm{Hz}^2$ respectively. Moreover, the vowel space of VC-1 and VC-2 appears to have similar shape to that of professional announcer. This result indicates that the converted voice might have better intelligibility than non-expert voice in noisy environment.
\subsubsection{eSTOI measurements}
\begin{figure}[bt]
    \centering
    \includegraphics[width=0.47\textwidth]{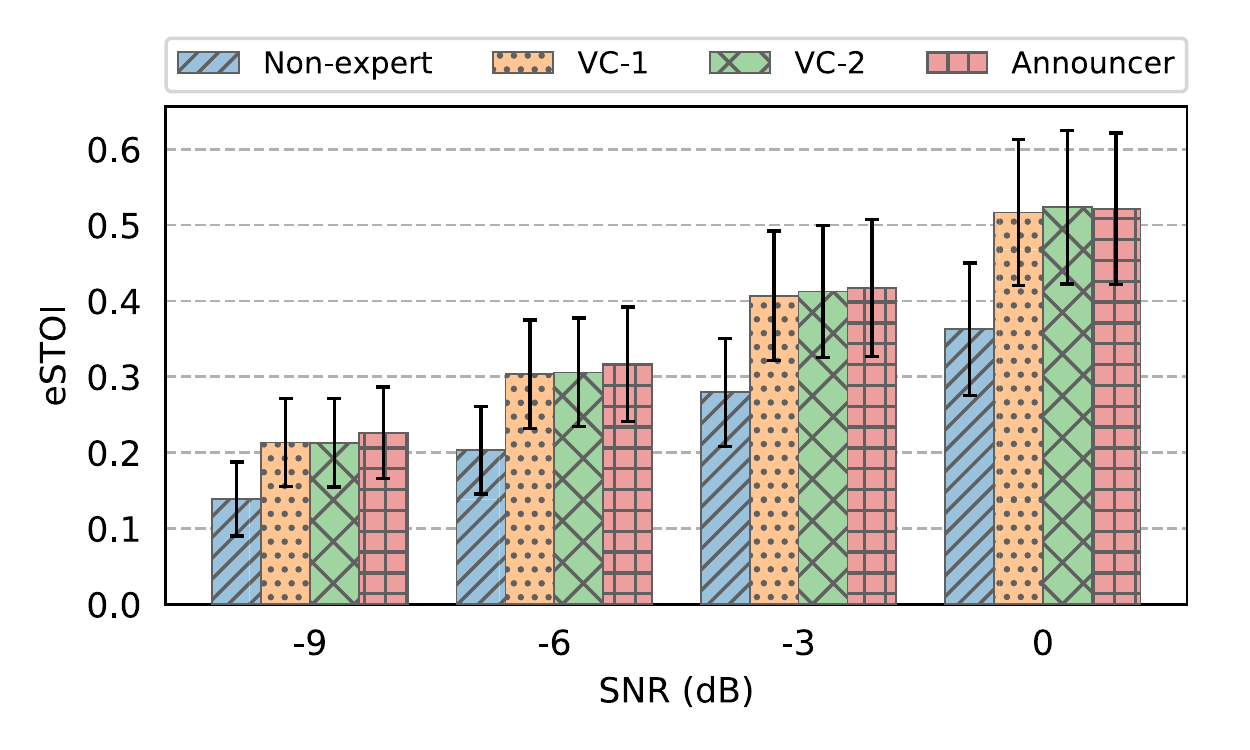}
    \caption{Average eSTOI score of non-expert, announcer, VC-1, and VC-2 stimuli across all utterances. The horizontal axis shows the SNR in dB. The eSTOI score is in range [0, 1], with higher score indicates better intelligibility.}
    \label{fig:estoi}
\end{figure}

To objectively measure the intelligibility of speech in noise, we calculate the eSTOI of the speech stimuli at 4 SNR levels: -9dB, -6dB, -3dB and 0dB using pySTOI python package\footnote{https://github.com/mpariente/pystoi}. The clean speech is used as the reference signal for eSTOI calculation. As can be seen from Fig. \ref{fig:estoi}, the announcer voice can resist to noisy environment better than non-expert voice as expected. Moreover, VC-1 and VC-2 stimuli also show a comparable performance to that of the announcer voice. This result confirms the effectiveness of our proposed method.

\begin{figure}
    \centering
    \includegraphics[width=0.47\textwidth]{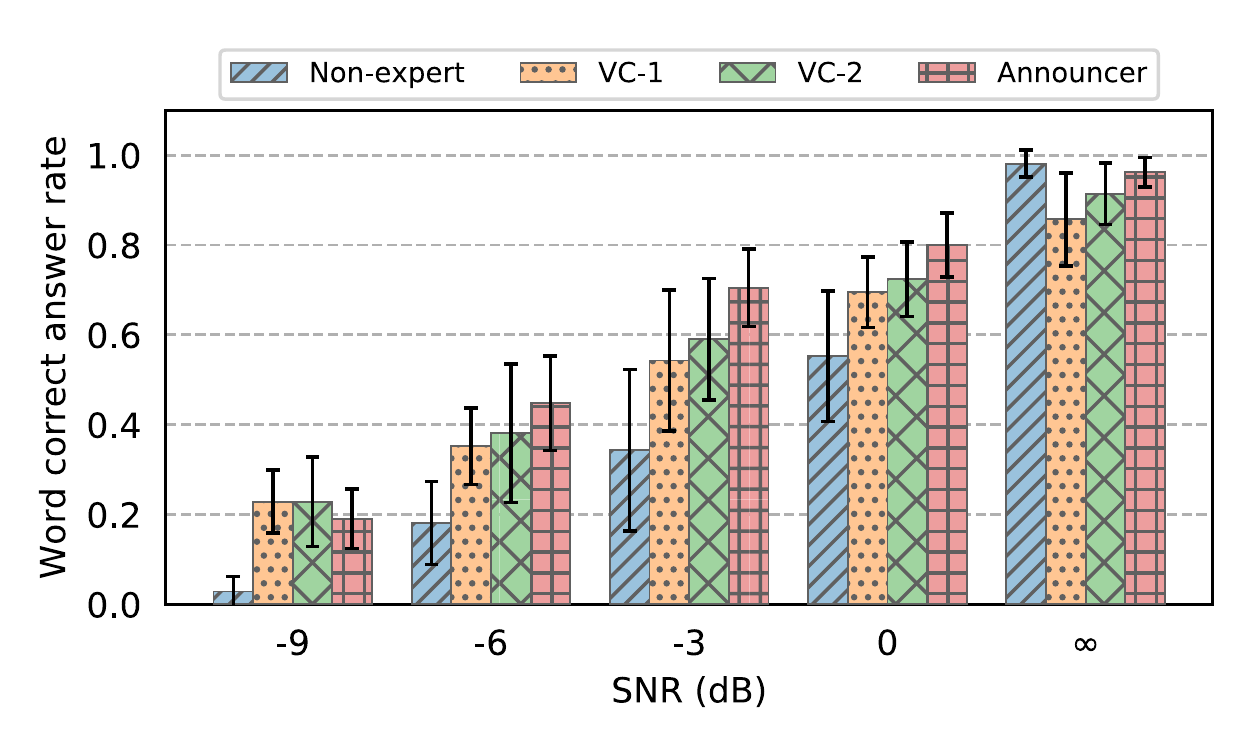}
    \caption{Average word correct answer rates and SE across participants. The horizontal axis shows the SNR in dB and $\infty$ denotes that the speech stimuli is presented without noise.}
    \label{fig:WRR}
\end{figure}
\begin{table*}[t]
\centering
\caption{\textit{p}-value of post-hoc pairwise comparison between 4 types of stimuli using Tukey’s HSD test. Values that reject the null hypothesis are in bold.}
\resizebox{\textwidth}{!}{
\begin{tabular}{ccccccc}
\toprule
SNR & Non-expert - VC1 & Non-expert - VC2 & Non-expert - Announcer & Announcer - VC1 & Announcer - VC2 & VC1 - VC2 \\
\midrule
$-9$ dB  & \textbf{0.001} & \textbf{0.001} & \textbf{0.003}  & 0.77  & 0.77 & 0.90 \\
$-6$ dB  & 0.064 & \textbf{0.024} & \textbf{0.0021} & 0.47  & 0.71 & 0.9  \\
$-3$ dB  & 0.1   & \textbf{0.031} & \textbf{0.001}  & 0.23  & 0.52 & 0.9  \\
$0$ dB   & 0.08  & \textbf{0.03}  & \textbf{0.001}  & 0.28  & 0.55 & 0.9  \\
$\infty$ & \textbf{0.016} & 0.32  & 0.9    & 0.051 & 0.59 & 0.45 \\
\bottomrule
\end{tabular}
}
\label{tab:p-value}
\end{table*}
\subsection{Subjective evaluations}
We conducted listening tests to compare the intelligibility of 4 types of speech stimuli. The listening test is conducted in sound-proof room to avoid any interference. Speech stimuli are presented via a D/A converter (RME, Fireface UCX), a headphone amplifier (STAX, SRM-1/MK2), and an electro-static headphone (STAX, SR-404). The sound pressure level is fixed at 60dB LAeq, measured on the clean speech stimuli by a sound level meter (Brüel\&Kjær, Types 2250) via an artificial ear (Brüel\&Kjær, Type 4153). Each participant listens to a set of 300 different random words, which are equally distributed into 5 SNR levels and 4 types of speech stimuli. The stimulus is presented only once for each trial and the order of presented speech stimuli is randomized for each participant. The duration of the whole listening test is approximately 40 minutes, which is divided into 4 sections with 2 minutes break between each section. There were 10 native Japanese participants, whose age ranging from 23 to 29 years old, joined our listening test. 

Before each test, the listeners are provided with instruction and some sample stimuli to get used to the sound level.
Figure \ref{fig:WRR} reports the average word correction rate across participants. One-way ANOVA test show that there were statistical differences between 4 types of stimuli.
A post-hoc pairwise analysis using Tukey HSD test ($p<0.05$) was carried out to determine statistical differences between pair of stimuli types in different SNR conditions. 

The results shown in Table \ref{tab:p-value} indicate that VC-2 and announcer stimuli are significant different from non-expert stimuli in noisy conditions ($\textrm{SNR} \leq 0\textrm{dB}$). In addition, no statistical difference between all 4 types of stimuli is found in clean condition ($\textrm{SNR}=\infty$). These results suggest that VC-2 is more effective than VC-1 for enhancing speech intelligibility in noisy condition. The possible reason for this difference might correspond to the increase amount of distortion when speaker individuality is completely changed. However, further analysis must be carried out to clarify this point.

\section{Conclusions}
The present work has proposed a speech intelligibility enhancement in noisy environment using voice conversion technique. The results from objective measurements and subjective evaluation confirm that adapting to announcer voice can increase the intelligibility of non-expert speaker. By analyzing the PCA of speaker embedding, it has been discovered that the announcer-speaking style is an independent features of speaker individuality. By modifying the second principal component of speaker embedding, we can manually control the amount of announcer-speaking style, hence increasing the intelligibility of speech in noisy environment without completely change the speaker individuality. Statistical analysis shows that modifying the second principal components yields the highest performance. Beside using announcer voice as the target, the proposed method can be applied to mimic Lombard speech and Clear speech, which are also speaking styles for increasing speech intelligibility in noisy condition. For the future work, the proposed method can be extended to generate converted speech adaptively to the noise condition in order to further improve the speech intelligibility in noise.

\section{Acknowledgements}
This work was supported by the SCOPE Program of Ministry of Internal Affairs and Communications (Grant number: 201605002).

\bibliographystyle{IEEEtran}
\bibliography{mybib}

\end{document}